\documentclass[aps,prl,twocolumn,floatfix,superscriptaddress,showpacs]{revtex4}
\usepackage{graphicx}% Include figure files
\usepackage{dcolumn}% Align table columns on decimal point
\usepackage{bm}% bold math
\usepackage{textcomp}

\begin{document}

\title{Gap opening in the zeroth Landau level of graphene}

\author{A.~J.~M.~Giesbers}
\affiliation{
High Field Magnet Laboratory, Institute for Molecules and Materials,
Radboud University Nijmegen, Toernooiveld 7, 6525 ED Nijmegen, The Netherlands
}

\author{L.~A.~Ponomarenko}
\affiliation{
Department of Physics, University of Manchester, M13 9PL, Manchester, UK
}

\author{K.~S.~Novoselov}
\affiliation{
Department of Physics, University of Manchester, M13 9PL, Manchester, UK
}

\author{A.~K.~Geim}
\affiliation{
Department of Physics, University of Manchester, M13 9PL, Manchester, UK
}
\author{M.~I.~Katsnelson}
\affiliation{Theory of Condensed Matter, Institute for Molecules
and Materials, Radboud University Nijmegen, Heyendaalseweg 135,
6525 AJ Nijmegen The Netherlands }

\author{J.~C.~Maan}
\affiliation{
High Field Magnet Laboratory, Institute for Molecules and Materials,
Radboud University Nijmegen, Toernooiveld 7, 6525 ED Nijmegen, The Netherlands
}

\author{U.~Zeitler}
\email[]{U.Zeitler@science.ru.nl}
\affiliation{
High Field Magnet Laboratory, Institute for Molecules and Materials,
Radboud University Nijmegen, Toernooiveld 7, 6525 ED Nijmegen, The Netherlands
}

\date{\today}% It is always \today, today,
% MAXIMUM 600 char including spaces
\begin{abstract}
We have measured a strong increase of the low-temperature resistivity
$\rho_{xx}$ and a zero-value plateau in the Hall conductivity $\sigma_{xy}$ 
at the charge neutrality point in graphene subjected to
high magnetic fields up to 30 T. 
We explain our results by a simple model
involving a field dependent splitting of the lowest Landau level of the
order of a few Kelvin, as extracted from activated transport measurements.
The model reproduces both the increase in $\rho_{xx}$ and the anomalous
$\nu=0$ plateau in $\sigma_{xy}$ in terms of coexisting electrons and 
holes in the same spin-split zero-energy Landau level.
 \end{abstract}

%%%%%%%%%%%%%%%%%%%%%%%%%%%%%%%%%%%%%%%%%%%%%%%%%%%%%%%%%%%%%%%%%%%%%%%%%%%%%%%%%%
\pacs{73.43.-f, % Quantum Hall effects, ...                                      %
      73.63.-b, %Electronic transport in nanoscale materials and structures      %
      71.70.Di} %Landau levels                                                   %
        %73.43.Qt, %Magnetoresistance                                            %
        %73.22.-f, %Electronic structure of nanoscale materials:                 %
%%%%%%%%%%%%%%%%%%%%%%%%%%%%%%%%%%%%%%%%%%%%%%%%%%%%%%%%%%%%%%%%%%%%%%%%%%%%%%%%%%

\maketitle

%Introduction 
In a magnetic field, graphene displays an unconventional Landau-level 
spectrum of massless chiral Dirac 
fermions~\cite{NovoselovNature, ZhangNature, Gusynin, McClure}.
In particular, a Landau level shared equally between electrons and holes 
of opposite chirality exists at zero-energy around the charge neutrality point (CNP).
Due to the coexistence of carriers with opposite charge, graphene
behaves as a compensated semimetal at the CNP with a finite 
resistivity $\rho_{xx}$ and a zero Hall resistivity $\rho_{xy}$.
 
Recently, the nature of the CNP in high magnetic fields has attracted  
considerable theoretical interest (see Ref.~\cite{reviewCastroNeto} 
and references there in).
Experimentally, in the metallic regime, the transport behavior around
the CNP can be explained using 
counter-propagating edge channels~\cite{Abanin_PRL}. On the other hand,
the high-field resistivity at the CNP was shown to diverge strongly, 
an effect recently analyzed in terms of a Kosterlitz-Thouless-type localization 
behavior~\cite{Checkelsky}.
In high quality graphene samples, made from Kish-graphite, 
Zhang et al.~\cite{Zhang_PRL, Jiang_PRL} have observed an
additional fine structure of the lowest Landau level 
in the form of a $\nu=\pm1$ state. The existence of this state
is proposed to be caused by a spontaneous symmetry breaking 
at the CNP and an interaction-induced splitting of the two levels
resulting from this.

Here we present an experimental study of the transport properties 
of the zero-energy Landau level in high magnetic fields
and at low temperatures. Calculating the conductivities from an
increasing magneto-resistance at the CNP and a zero-crossing of the
Hall resistance yields a zero minimum in the longitudinal conductivity
$\sigma_{xx}$ and a quantized zero-plateau in the Hall-conductivity
$\sigma_{xy}$. The temperature dependence of the $\sigma_{xx}$-minimum 
displays an activated behavior. We explain this transition with a simple model involving 
the opening of a spin-gap (30 K at 30 T) in the zeroth Landau level. 
We do not observe any indication for
a spontaneous symmetry breaking and an interaction-induced splitting at $\nu=\pm 1$
as reported in Refs.~\onlinecite{Zhang_PRL} and  ~\onlinecite{Jiang_PRL}
and we tentatively assign this to the relatively larger disorder in our samples
made from natural graphite.

\begin{figure}[tb]
  \begin{center}
  \includegraphics[width=0.9\linewidth]{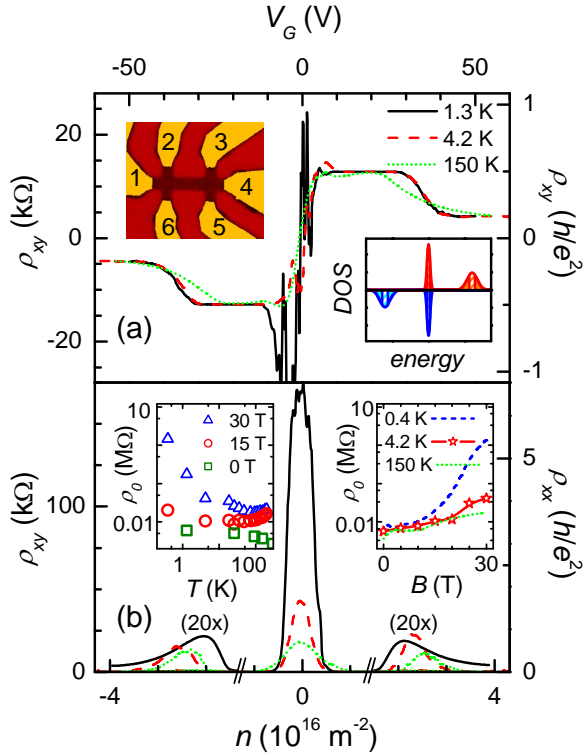}
  \end{center}
\vspace*{-1em}
\caption{ (color online) (a) $\rho_{xy}$ as a function of
back-gate voltage $V_{G}$ (top $x$-axis) and charge carrier concentration $n$ 
(bottom $x$-axis) measured between contacts 3 and 5 as shown in the scanning 
electron micrograph in the top left inset at a fixed magnetic field $B=30$~T.
(b) $\rho_{xx}$ at $B=30$~T measured at the same temperatures 
as in (a). Note the x-axis break and the 20-times magnification 
for high electron/hole concentrations.  
The left inset in (b) shows the temperature dependence 
of $\rho_{xx}$ at the charge CNP for different $B$. The right inset displays 
the magnetic field dependence of $\rho_{xx}$ at the CNP for different $T$.
}
\label{CNP}
\end{figure}

The monolayer graphene devices (see top left inset Fig.~\ref{CNP}a) 
are deposited on Si/SiO$_2$ substrate using methods as already reported 
elsewhere~\cite{NovoselovScience, AndreNatMat}. The doped Si acts as a back-gate
and allows to adjust the charge-carrier concentration in the graphene film
from highly hole-doped to highly electron-doped. 
Prior to the experiments the devices were annealed at 390 K during several 
hours removing most of the surface impurities~\cite{SchedinNatMat} and 
increasing the low-temperature mobility from  
$\mu=0.47$~m$^2$(Vs)$^{-1}$ to $\mu=1.0$~m$^2$(Vs)$^{-1}$.
All measurements were preformed with standard AC-techniques at low enough 
currents to avoid any heating effects.

%Broader introduction, half integer QHE, Figs. 1a and 1b
Charge carriers in graphene behave as massless chiral Dirac
fermions with a constant velocity $c \approx 10^{6}$~m/s and thus
a linear dispersion $E= \pm c \hbar
|k|$~\cite{Wallace,NovoselovNature,ZhangNature}. The $\pm$ sign
refers to electrons with positive energies and holes with negative
energies, respectively. In a magnetic field, this linear spectrum
splits up into non-equidistant Landau levels at $E_N = \pm
c\sqrt{2e\hbar BN}$~\cite{McClure}. Higher Landau levels ($N \ge
1)$ are fourfold degenerate and filled with either electrons
($E>0$) or holes ($E<0$). The zeroth Landau level ($N=0$) is half
filled with electrons and half filled with holes of opposite
chirality.

A consequence of this peculiar Landau-level structure is the
half-integer quantum Hall effect
\cite{NovoselovNature,ZhangNature}. Figure~\ref{CNP}a and b
visualize this effect by means of magneto-transport experiments in
a graphene field-effect transistor (FET) at $B=30$~T. At large
negative back-gate voltages ($V_g < -40$~V) two hole levels are
completely filled, the Hall resistivity is quantized to
$\rho_{xy}= h/6e^{2} = 4.3~$k$\Omega$ and the longitudinal
resistivity $\rho_{xx}$ develops a zero-minimum. Moving $V_g$
toward zero depopulates the $N=1$ hole level and moves the Hall
resistance to the following plateau, $\rho_{xy}= h/2e^{2} =
12.9~$k$\Omega$ accompanied by another zero in $\rho_{xx}$.
Sweeping $V_g$ through zero then depopulates the hole states in
the zeroth Landau level and, simultaneously, populates electron states in the same
level. $\rho_{xy}$ changes smoothly through zero
from its negative quantized value on the hole side to a positive
quantized value on the electron side whereas $\rho_{xx}$ moves
from a zero on the hole side through a maximum at the CNP to
another zero-minimum on the electron side.

%%%%%%%%%%%%%%%%%%%%%%%%% figure 1 %%%%%%%%%%%%%%%%%%%%%
%%%%%%%%%%%%%%%%%%%%%%% End figure 1 %%%%%%%%%%%%%%%%%%%

In the following we will concentrate on the electronic behavior
of graphene around the CNP in high magnetic fields and at low
temperatures. Our experimental findings are summarized 
in Figure~\ref{CNP}c. For high magnetic fields
($B>20$~T) $\rho_{xx}$ strongly increases from a few tens of  
kilo-ohms to several mega-ohms when the temperatures is lowered down to 0.4 K
(left inset in Fig.~\ref{CNP}c). Such a field-induced transition
can also be seen in the right inset in
Fig.~\ref{CNP}c: Whereas the magnetoresistance displays a moderate
field dependence for sufficiently high temperatures, it starts to
increase strongly for $B>20$~T for the lowest temperature $T =
0.4$~K.

At zero magnetic field the resistivity at the CNP behaves as 
can be expected, it increases with decreasing temperature toward
$h/4e^{2}$, as carriers are slightly frozen out.

To gain more insight into the peculiar behavior at the CNP we
have calculated the longitudinal conductivity $\sigma_{xx}$ and
Hall conductivity $\sigma_{xy}$ from the experimentally measured
$\rho_{xx}$ and $\rho_{xy}$ by tensor inversion. Possible
artifacts due to admixtures of $\rho_{xx}$ on $\rho_{xy}$ and vice
versa, caused by contact misalignment and inhomogeneities, were
removed, by symmetrizing all measured resistances for positive and
negative magnetic fields.

\begin{figure}[tb]
  \begin{center}
  \includegraphics[width=0.9\linewidth,angle=0]{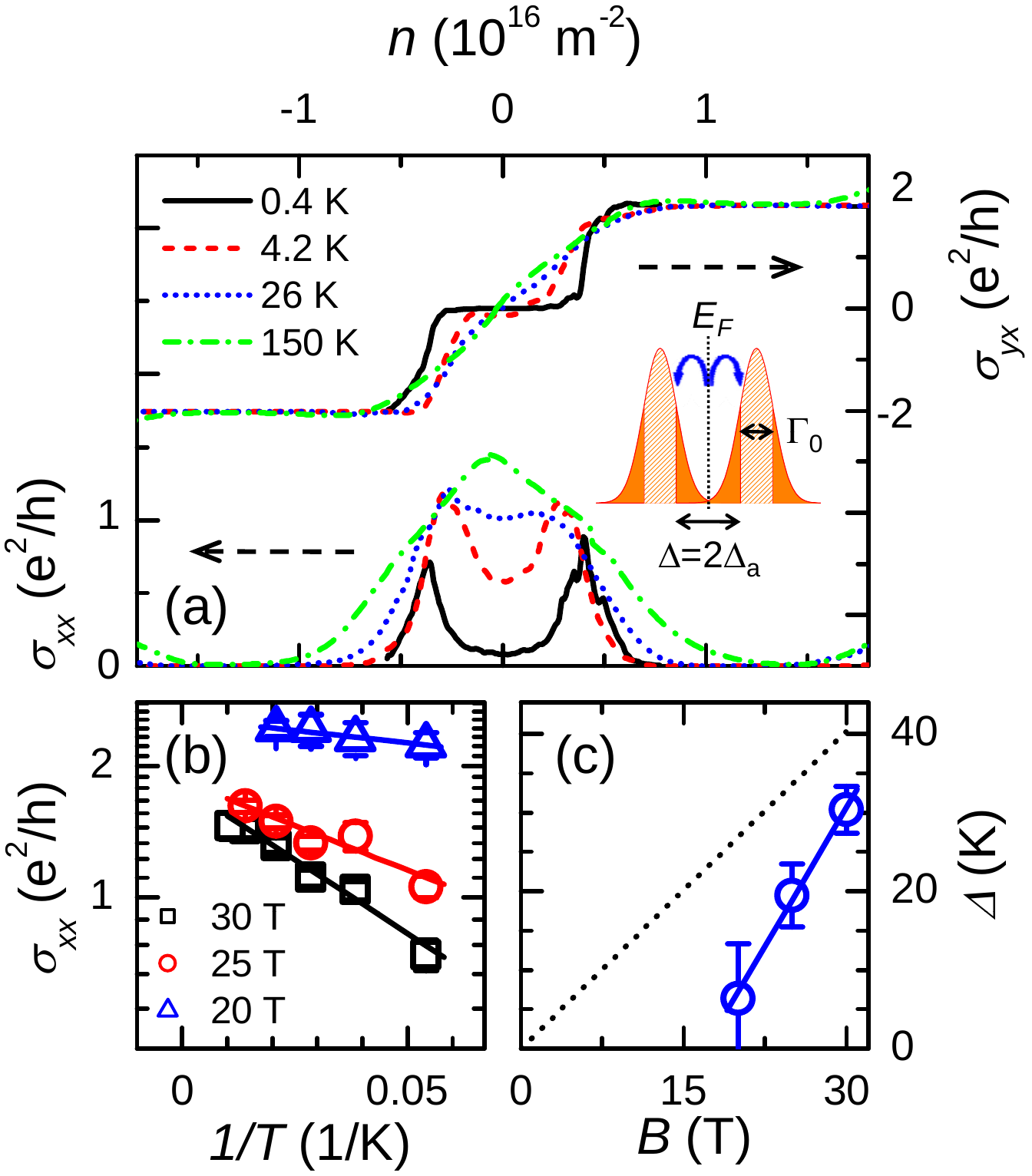}
  \end{center}
  \vspace*{-1em}
  \caption{ (color online) (a) Concentration dependence of $\sigma_{xy}$ 
  (top panel) and $\sigma_{xx}$ (bottom panel) at $B=30$~T. 
  The inset sketches the principle of activated transport between two 
  broadened Landau levels. 
  (b) Arrhenius plots of the temperature dependence of $\sigma_{xx}$ at 
  the CNP in the thermally activated transport regime.
  (c) Energy gaps $\Delta$ as extracted from the slopes 
  of the Arrhenius plots. For comparison, the dotted black line represents 
  the expected gap for a bare Zeeman spin-splitting $\Delta_Z = g \mu_{B} B$ 
  with a free-electron $g$-factor $g=2$.}
  \label{Sigmas}
\end{figure}

Fig.~\ref{Sigmas}a shows the two components of the conductivity
tensor at $B =30$~T: At low temperatures, the Hall conductivity
$\sigma_{xy}$ develops an additional, clearly pronounced,
zero-value quantum Hall plateau around the CNP and the
longitudinal conductivity $\sigma_{xx}$ displays a thermally
activated minimum. Both features are an immediate consequence of
the strongly increasing $\rho_{xx}$ at the CNP.

%%%%%%%%%%%%%%%%%%%%%%%% figure 2 %%%%%%%%%%%%%%%%%%%%%%%
%%%%%%%%%%%%%%%%%%%%%%%%% End figure 2 %%%%%%%%%%%%%%%%%%%

Both the minimum in $\sigma_{xx}$ and the
zero-value plateau in $\sigma_{xy}$ at the CNP disappear at
elevated temperatures (Fig.~\ref{Sigmas}a) pointing toward an
activated transport behavior. Indeed, when plotting $\sigma_{xx}$
in an Arrhenius plot (Fig.~\ref{Sigmas}b),
\begin{equation}\label{gap}
  \sigma_{xx} \propto \exp{(-\Delta_{a}/kT)},
\end{equation}
we can reasonably fit the data with a 
single Arrhenius exponent, yielding an activation gap $\Delta_{a}$.
For low temperatures, we do no more observe a simple activated behavior since
other transport mechanisms such as variable range
hopping~\cite{VRH1,VRH2} start to play an important role.
For the lowest temperatures ($T<1$~K), the resistance starts to diverge strongly,
an effect which has been suggested to be caused by the abrupt transition
to an ordered truly insulating state~\cite{Checkelsky}
and comes on top of the simple activated behavior.

We can relate the observed activated behavior
to the opening of a gap in the zeroth Landau level 
(inset in Fig.~\ref{Sigmas}a), 
$\Delta = 2 \Delta_a$, and, therefore, to a partial lifting of its four-fold
degeneracy. The experimentally determined field dependence of this
gap is shown in Fig.~\ref{Sigmas}c. The gap only
becomes visible for fields above 15 T because of a finite width
$\Gamma_0$ of the split Landau level at zero energy; linear extrapolation
of the experimental data to zero field yields $\Gamma_0 \approx
30$~K. The linear variation of $\Delta$ with magnetic field
suggests that the splitting might be spin related. For
comparison we have plotted the theoretical Zeeman splitting
$\Delta_Z = g\mu_{B} B$ for a spin-split zeroth Landau level,
with a free-electron $g$-factor $g =2$~\cite{Zhang_PRL}
($\mu_B$ the Bohr magneton). The slightly higher slope of the
experimentally measured gap may be explained by a sharpening of 
the zeroth Landau level (i.e. a reduction of  $\Gamma_0$)
with increasing field~\cite{Giesbers_PRL}, or, alternatively by 
a slight exchange enhancement of the splitting. 

%%%%%%%%%%%%%%%%%%%%%%%%% figure 3 %%%%%%%%%%%%%%%%%%%%%%%
\begin{figure}[tb]
  \begin{center}
  \includegraphics[width=0.9\linewidth,angle=0]{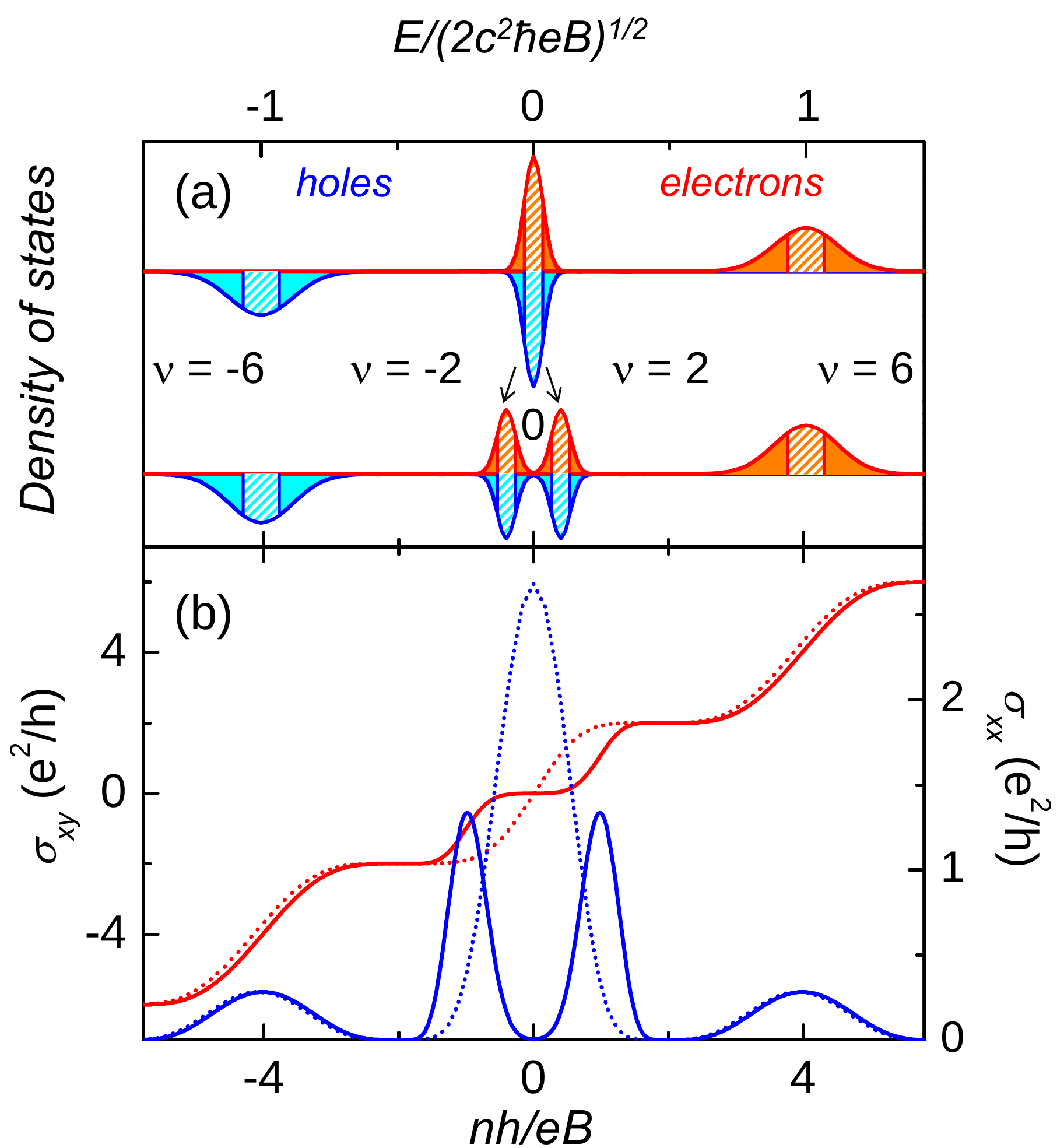}
  \end{center}
  \vspace*{-1em}
\caption{(a) Schematic illustration of the density 
of states for electrons (red) and holes (blue) in graphene in a 
magnetic field without spin-splitting (top) and including spin splitting
(bottom). The width of the higher Landau levels is in reality much larger
than illustrated (see text).
(b) $\sigma_{xy}$ and  $\sigma_{xx}$ as a function 
of charge carrier concentration for unsplit (dotted lines) and 
split (solid line) Landau levels calculated for $B=30$~T and $T=1$~K
and using $g=2$.
}
\label{Model}
\end{figure}
%%%%%%%%%%%%%%%%%%%%%%%%% End figure 3 %%%%%%%%%%%%%%%%%%%

%Model
The consequences of the opening of a gap can be understood
intuitively using a simple model (Fig.~\ref{Model}): The density
of states of graphene in a magnetic field consist of a series of
Landau levels occupied by electrons or holes. Each
Landau level is represented by a Gaussian with a degeneracy of
$2eB/h$ for the individual electron and hole state of the zeroth
Landau level and $4eB/h$ for the higher levels. The zeroth level
is relatively sharp~\cite{Giesbers_PRL} whereas the
higher levels are considerably broadened by, amongst others, 
random fluctuations of the perpendicular magnetic field
caused by the rippled surface of graphene~\cite{rippled-surface1,
rippled-surface2, rippled-surface3}. Electrons and holes in the
center of the Landau levels are extended (shaded areas) whereas 
they are localized in the Landau level tails (filled areas). 

With this simple model-density-of-states one can straightforwardly calculate
the longitudinal conductivity $\sigma_{xx}$ (Fig.~\ref{Model}b) by
means of a Kubo-Greenwood formalism~\cite{Kubo, Greenwood} and the
Hall conductivity $\sigma_{xy}$ (Fig.~\ref{Model}b) by summing up
all the states below the Fermi-energy~\cite{Streda}. The results
are shown by the dotted lines in Fig.~\ref{Model}b and 
reproduce all the features of a half-integer quantum Hall effect
in graphene~\cite{NovoselovNature, ZhangNature}.

An additional spin-splitting of all the Landau levels is schematically
introduced into the model in the lower part of Fig.~\ref{Model}a. 
The splitting is only
resolved in the lowest Landau level, where the level width is
already small enough (20 K). Due to the large broadening of the higher 
levels (400 K~\cite{Giesbers_PRL}) it remains unresolved here. 
Fig.~\ref{Model}b shows how this 
splitting nicely reproduces the plateau in $\sigma_{xy}$
accompanied by a minimum in $\sigma_{xx}$. As soon as the
splitting exceeds the width of the extended states in
the lowest Landau level, the resistivity $\rho_{xx}$ at the CNP 
starts to increase as a natural consequence of the 
tensor inversion of the developing minimum in $\sigma_{xx}$ and 
the quantized Hall plateau in $\sigma_{xy}$.

%Discussion: electrons and holes

The proposed density of states in Fig.~\ref{Model}b. not only
explains the experimentally observed increase in $\rho_{xx}$ at the CNP but also the
behavior of  $\rho_{xy}$. In conventional semiconductors with a gap,
$\rho_{xy}$ develops a singularity around around the CNP when either electrons
or holes are fully depleted. In our graphene samples, however, $\rho_{xy}$
is moving through zero from the $\nu=-2$ plateau with
$\rho_{xy}=-h/2e^2 = 12.9$~k$\Omega$ to the $\nu=+2$ plateau at
$h/2e^2$. This observation implies that electrons and holes are
coexisting both below and above the CNP and graphene behaves as a
compensated semimetal with as many hole states as electron
states occupied at the CNP.

It should be mentioned in this context that
Zhang \textit{et al.} have observed an additional splitting of the
lowest Landau level in other samples~\cite{Zhang_PRL, Jiang_PRL} resulting
into additional quantum Hall plateaus at 
$\rho_{xy} = \pm h/e^2 = \pm 25.8$~k$\Omega$ 
and $\nu=\pm 1$ minima in $\rho_{xx}$. 
To explain these results, 
the degeneracy of electron and hole states has to be removed by
lifting the sublattice degeneracy. 
In this respect, the absence of $\nu=\pm 1$ state in our and 
other experiments~\cite{Abanin_PRL, Checkelsky, Giesbers_PRL} 
points to a larger disorder, confirmed by the lower mobility,
which prevents symmetry breaking. Therefore, we deal with a symmetric 
density of states for electrons and holes with two two-fold 
degenerate spin-split levels shared by electrons and holes simultaneously. 

An important consequence of our experiments
is that the gaps measured are more than an order of magnitude
smaller than interaction induced gaps. Indeed, the
characteristic energy of electron-electron correlation $I =
e^2/4 \pi \epsilon_{0} \epsilon_{r} l_{B}$ (where $l_{B}=\sqrt{\hbar/eB}$ 
is the magnetic length and $\epsilon_{r} \approx 2.5$ is the 
effective dielectric constant for graphene on silicon
dioxide) amounts to about 1400 K at 30 T and exceeds the width 
of the zero-energy Landau level $\Gamma_{0}$ by nearly two orders of 
magnitude. According to a simple Stoner-like theory of quantum Hall
ferromagnetism~\cite{nomura} this should result in spontaneous
breaking of spin and, possibly, valley degeneracy with the
formation of quantum Hall ferromagnetism. Since we do not see any
evidences of the spontaneous breaking, the effective Stoner parameter is
apparently reduced to the value of order of $\Gamma_{0}$. Probably 
the sample mobility is not high enough for the stable appearance of
a ferromagnetic phase as proposed in Fig.~1 of Ref.~\onlinecite{nomura}.
A suppression of the Stoner parameter is also what
one would naturally expect for narrow-band itinerant-electron 
ferromagnets due to so called $T$-matrix renormalization~\cite{kanamori, edwards}.
Unfortunately, for the case of quantum-Hall ferromagnetism a
consequent theoretical treatment of the effective Stoner criterion
with a controllable small parameter is only possible for the limit
of high Landau levels~\cite{glazman, fogler} and, admittedly, numerous
other theoretical scenarios may be considered (see e.g.
Ref.~\onlinecite{reviewCastroNeto} and references therein). 
Therefore, further theoretical efforts are certainly
required to clarify this issue.

% Conclusion
In conclusion, we have measured the transport properties of the zeroth Landau level in
graphene at low temperatures and high magnetic fields.
An increasing longitudinal resistance accompanied by a zero crossing in the
Hall resistivity at the CNP leads to a flat plateau in the Hall conductivity together
with a thermally activated minimum in the longitudinal conductivity.
This behavior can be related to the opening of a Zeeman-gap in the density of states of the
zeroth Landau level, which reasonably reproduces the features observed experimentally.

This work is part of the research program of the 
'Stichting voor Fundamenteel Onderzoek der Materie (FOM)', 
which is financially supported by the 'Nederlandse Organisatie voor Wetenschappelijk Onderzoek (NWO)'.
It has been supported by EuroMagNET under EU contract RII3-CT-2004-506239, 
the EPSRC, the Royal Society and the European Research Council (Program 'Ideas', Call:
ERC-2007-StG).

%%%%%%%%%%%%%%%%%%%%%%%%%%%%%%%%%%%%%%%%%%%%%%%%%%%%%%%%%%%%%%%%%%%%%%%%%%%%%%%%

\end{document}